\documentclass[12pt]{iopart}

\usepackage{iopams}
\ifx\pdfoutput\undefined
   \usepackage[dvips]{graphicx}
   \else
   \usepackage[pdftex]{graphicx}

   \usepackage{epstopdf}

\begin{document}

\title[NIR transmission spectrum of the Hot-Jupiter XO-1b]{Probing the terminator region atmosphere of the Hot-Jupiter XO-1b with transmission spectroscopy}

\author{G. Tinetti$^1$,
P. Deroo$^{2}$,
M.~R. Swain$^{2}$,
C. ~A.~Griffith$^{3}$,
G. Vasisht$^{2}$, 
L.~R. Brown$^{2}$,
C. Burke$^{4}$, \&
P. McCullough$^{5}$}

\address{  $^1$University College London, Dep. of Physics and Astronomy, Gower Street, London WC1E 6BT, UK; \newline
$^2$Jet Propulsion Laboratory, California Institute of
  Technology, 4800 Oak Grove Drive, Pasadena, CA 91109; \newline
$^3$University of Arizona, Lunar and Planetary
  Laboratory, 1629 E. University
  Bvld. Tucson, AZ 85721; \newline
$^4$ Harvard-Smithsonian Center for Astrophysics, 60 Garden St.,
MS-65, Cambridge, MA 02138;         \newline
$^5$ Space Telescope Science Institute, 3700 San Martin Dr., Baltimore, MD 21218}

\ead{g.tinetti@ucl.ac.uk}

\begin{abstract}
{We report here the first infrared spectrum of the
  hot-Jupiter XO-1b. The observations were obtained with NICMOS instrument onboard the Hubble Space
  Telescope during a primary eclipse of the
  XO-1 system.  Near photon-noise-limited spectroscopy between 1.2 and
  1.8 $\mu$m allows us to determine the main composition of this
  hot-Jupiter's planetary atmosphere with good precision.  This is the third hot-Jupiter's atmosphere for which spectroscopic data are available in the near IR.
  The spectrum
  shows the presence of water vapor (H$_{2}$O), methane (CH$_{4}$) and
  carbon dioxide (CO$_2$), and suggests the possible presence of
  carbon monoxide (CO). 
  We show that the published IRAC secondary transit emission photometric data are compatible
  with the atmospheric composition at the terminator determined from the NICMOS spectrum, with a range of possible
  mixing-ratios and thermal profiles;  additional emission spectroscopy data
  are needed to reduce the degeneracy of the possible
  solutions. Finally, we note the similarity between the 1.2-1.8 $\mu$m transmission spectra of XO-1b
  and HD 209458b, suggesting that in addition to having similar
  stellar/orbital and planetary parameters the two systems may also have a similar exoplanetary atmospheric composition. }
\end{abstract}

\maketitle

\section{Introduction}
\label{sec:intro}
The exoplanet XO-1b is the first one discovered by the XO project
(McCullough et al., 2006), and belongs to the so called hot-Jupiter
class of objects, i.e. gas giant planets orbiting very close to their
parent stars. XO-1 is a $V=11.3$ G1V star, almost identical to our own
Sun. 
The primary transit and radial velocity measurements (McCullough et al., 2006; Holman et al., 2006; Torres et al., 2008)
are consistent with a planet of mass  $M_{\rm p} = 0.9 (\pm 0.07) M_{\rm J}$, planetary/stellar radii ratio $R_{\rm p}/R_*=  0.1326\pm0.0004$,
 semi major axis $a = 0.04928\pm0.00089$ AU and a stellar $T_{\rm eff}\sim 5750$ K.
The equilibrium temperature of XO-1b is $\sim 1200$ K.
Follow-up observations of XO-1 with the IRAC camera on board the  Spitzer Space Telescope
were recently obtained (Machalek et al., 2008), providing new insight into the thermal properties of this exoplanet atmosphere.

We present here new primary transit spectroscopy data of the exoplanet
XO-1b obtained with the NICMOS instrument on board the Hubble Space
Telescope.  A combination of high quality data and the presence of
strong molecular features offers the opportunity for a robust determination of
chemical composition of a hot-Jupiter's atmosphere.
Having secured the atmospheric composition of XO-1b with the NICMOS transmission spectrum  in the
terminator region, the emission photometry data recorded by Machalek et al. (2008) can be used to constrain the vertical thermal structure of the planetary atmosphere on the day side. Emission data are, in fact,  highly sensitive to atmospheric thermal gradients and thus  complement 
transmission spectra, which are less dependent on the temperature profile.

We remark that the stellar/planetary and orbital parameters of  XO-1b are quite similar to the ones of the hot-Jupiter HD
209458b (Mazeh et al., 1999; Charbonneau et al., 2000, Torres et al., 2008), 
for which
$M_{\rm p} = 0.685\pm0.015 M_{\rm J}$, 
$R_{\rm p}/R_* =  0.12086\pm0.0001$,
 semi major axis $a =  0.04707\pm0.00046$ AU,  stellar $T_{\rm eff} \sim 5942$ K.
The equilibrium temperature of HD 209458b is $\sim 1450$ K; photometric and spectroscopic data in the NIR and MIR were used to constrain
its thermal profile (Swain et al., 2009b).

Previous observations of HD 209458b   with
NICMOS were made in the same instrument configuration of the spectrum reported here (Deroo et al., submitted) thus providing an ideal basis for comparison.
While H$_{2}$O, CH$_{4}$ and CO$_2$ are present in both HD 209458b and XO-1b, the HD 209458b spectrum suggests the additional presence of hazes,
non detectable in the atmosphere of XO-1b.

\section{Observations \& Data Analysis}
\label{ab}
The transit of XO-1b in front of its host star was monitored spectro-photometrically using HST/NICMOS on February 9 and 21, 2008. The observations were performed using the G141 grism covering the wavelength range between 1.2 and 1.8 $\mu$m. The instrument was defocused resulting in a spectral resolution of $R \sim 35$. In both cases, five HST orbits cover the event.  This paper is focused on the second observation for two reasons: (1) the second observation covers the full eclipse in two HST orbits instead of one for the first observation, and (2) the raw signal of the first observation is less stable than the second observation by a factor 2.5.
 As a result, even though both transmission spectra are compatible, only the second observation provides sufficient signal to noise for molecular spectroscopy. An upcoming publication (Burke et al., in prep)  will focus on the ephemerides derived from these measurements and for this, the first observation is the most decisive dataset.
To combine the two observations one would need to assume that the distribution of stellar spots remained unchanged, but such an assumption is not granted. First of all stellar spots are not uniformly distributed in longitude, as observed in the Sun and as inferred by the light curve modulation in other stars, as e.g. observed by CoRoT (Mosser et al. 2009). The low projected rotational velocity of the XO-1 star implies a period comparable to the solar one ($v \sin i \leq 3$ km s$^{-1}$, McCullough et al. 2006): since our two observations are separated by ca. 12 days, the XO-1 star will have shown two different hemispheres on the two dates, with likely two different spot distributions. Additionally, as shown by Mosser et al. (2009), the stellar spot distribution evolves on time scales comparable to the rotational period of the star. Therefore, even if  the transit observations were planned to fall always against the same stellar hemisphere, still the spot distribution would be different. As a consequence, primary eclipse average depths can and do change due to variable star spot distribution, and there is no obvious way to correct for them at the level of accuracy required for molecular spectroscopy.

The observation on February 21, 2008 covers the pre-, full and post-transit phase with two, two, and one HST orbit respectively. In total, 279 spectra were measured with an integration time of 40 sec each. Because the signal was still stabilizing in the first orbit, we omitted this data in the analysis. We analyzed the observations with a previously established method (Swain et al., 2008b, Swain et al., 2009a, Swain et al., 2009b). The recorded signal needs to be corrected for systematics, which is done through a multi-dimensional minimization taking into account the planet's orbital parameters (transit depth, semi major axis, inclination, eclipse timing etc.) previously obtained (McCullough et al.,  2006; Holman et al., 2006; Torres et al., 2008) and  the optical state vectors 
 (see Swain et al. 2008b). 
Limb darkening parameters were taken from Claret (2000) for the $J$ and $K$ bands and interpolated to the appropriate wavelength. In the minimization, the only free parameter is the depth of the eclipse, which is derived for each spectral channel. The final spectrum  is shown in Fig. 1. 

The zero point of the derived spectrum implies $(R_{\rm p}/R_*)^2 = 1.71 \pm 0.02 $ \%, which is within 1-$\sigma$ from the radius ratio reported in 
McCullough et al. (2006) and Holman et al. (2006).  The error bar on the zero point is significantly larger than the point-to-point uncertainty on the spectrum, since it is more sensitive to errors introduced   by stellar variability or limits in the ephemeris determination. 
For this reason, the approach based on multiple bands observed at different times, recently proposed by Sing et al., (2009) as an alternative to transmission spectroscopy, 
cannot provide the level of accuracy needed to detect molecular signatures.
In contrast, simultaneous measurements at all relevant wavelengths 
 make the measured spectral shape  
 relatively insensitive to such errors and quite robust  to determine the presence of molecular species in the 1.2-1.8 $\mu$m region.

\section{Spectral interpretation}

We modeled the transmission spectrum of XO-1b using line by line
radiative transfer models optimized for hot-Jupiter planets, which account for the effects of molecular opacities (Tinetti et al.,
2007a,b) and hazes (Griffith et al. 1998).  
In our simulations we included H$_{2}$O, CH$_{4}$, CO and CO$_2$.
While the  BT2 line list for water (Barber et
al., 2006) can be calculated at the appropriate temperatures,  
the available data lists for methane at high temperature are inadequate to probe the  modulations of the atmospheric
thermal profile. 
To cover the spectral range of our data, we had to use
multiple data lists for methane, HITRAN 2008 (Rothman et al., 2009),
PNNL, and hot-temperature measurements at 800, 1000 and
1273 K (Nassar and Bernath, 2003).   For CO$_2$ we used HITEMP (Rothman et al.,
private comm.) and CDSD-1000 (Tashkun et al., 2003), for CO we
used HITEMP.  The contribution of H$_2$-H$_2$ at high
temperatures was taken from Borysow et al. (2001).

\begin{figure}
\includegraphics{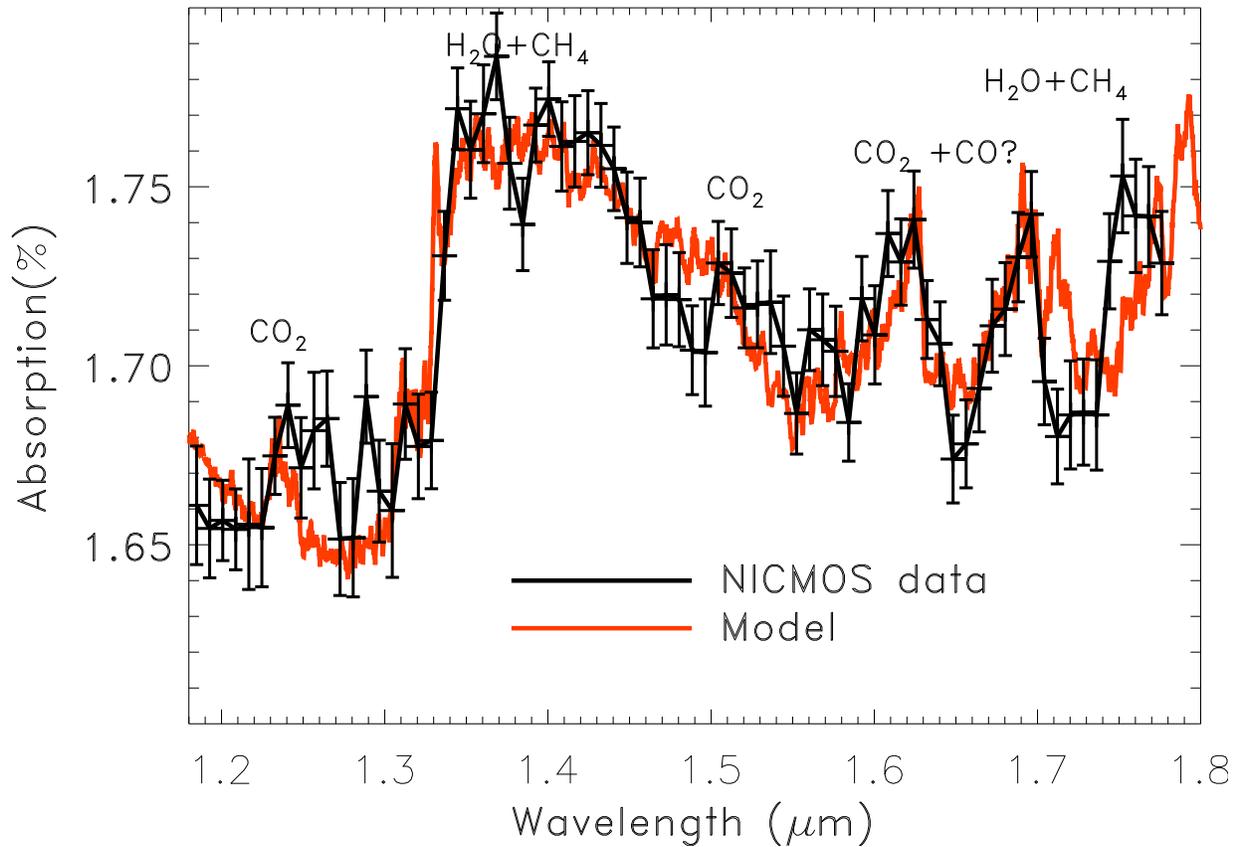}
\caption{The primary transit spectrum of XO-1b obtained with the Hubble-NICMOS instrument. The atmospheric model containing H$_{2}$O, CH$_{4}$, CO and CO$_2$ is shown by the orange line. While we have no doubts concerning the presence of  H$_2$O, CH$_4$  and CO$_2$ in the exoplanet's atmosphere, the additional presence of CO is 
less secure, due to the poor knowledge of the absorption  of methane at high temperature in this spectral region. }
\label{xo2}
\end{figure}

\begin{figure}
\includegraphics{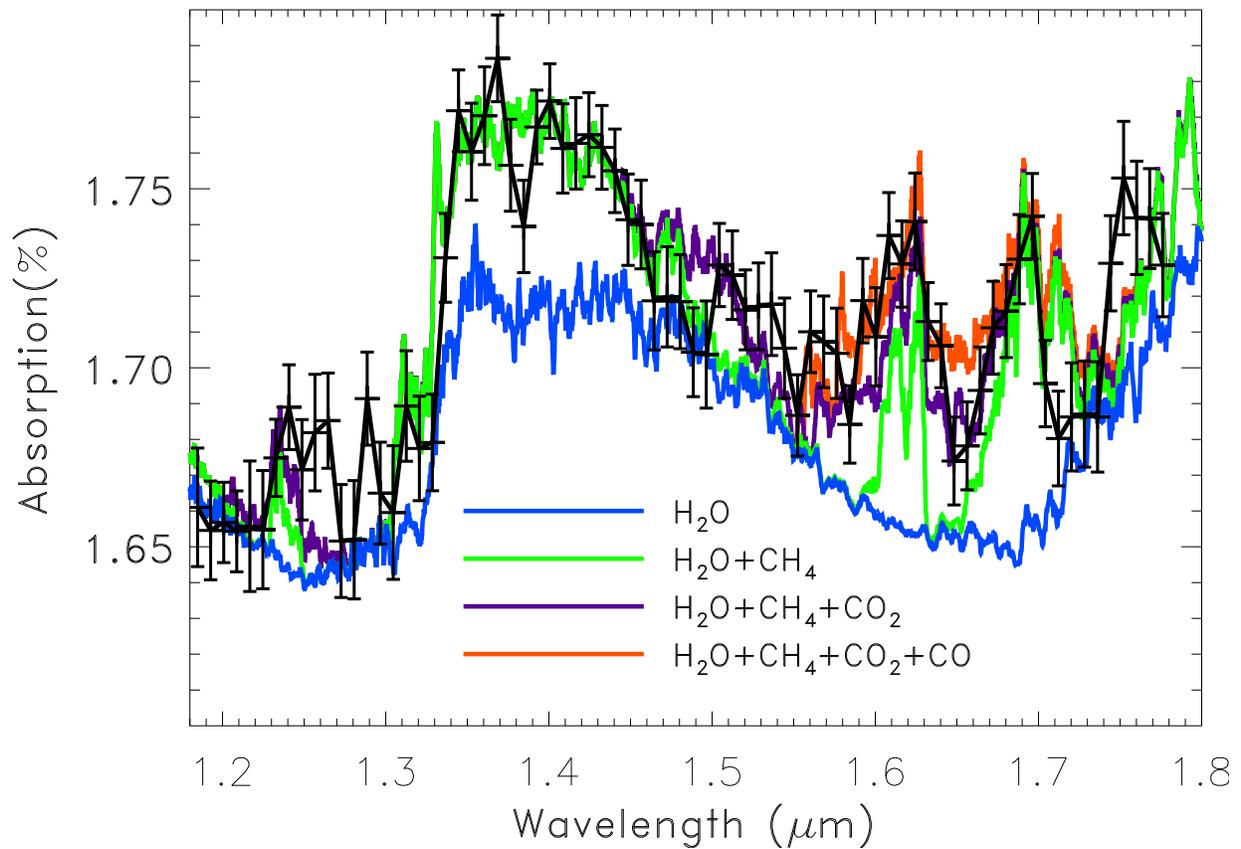}
\caption{Same spectrum as in Fig. 1, with the contribution of the different molecules present in the atmosphere shown separately.}
\label{mol}
\end{figure}

We find that absorption due to H$_{2}$O, CH$_{4}$  and CO$_2$  explains all the features visible in the NICMOS spectrum. The additional contribution
of CO may improve the fit between 1.55 and 1.75 $\mu$m, Fig. \ref{xo2} and Fig. \ref{mol}, but we cannot discard the possibility that improved data lists for methane 
and/or CO$_2$ would provide the missing opacity. 

The H$_{2}$O abundance determined from the spectrum depends on the assumed temperature and planetary/stellar radii. Although transmission spectra are not very sensitive to thermal
gradients, certainly not as sensitive as emission spectra, the atmospheric
temperature may play an important role in the overall scale height,
hence in the amplitude of the spectral signatures, as well as in the
molecular absorption coefficients. A range of $T-P$ profiles similar to those plotted in Fig. 3 combined with a thermochemical equilibrium H$_{2}$O 
abundance of $4.5 \times 10^{-4}$ 
(Liang et al., 2003; 2004) provide an excellent match. However, a
$\sim$ 1\% difference in the estimate of the planetary radius at the $\sim$ 1 bar pressure level, 
would result in a variation of the H$_{2}$O abundances of a factor of 10. Additional primary 
transit data at different wavelengths are needed to  constrain the H$_{2}$O abundance. The mixing ratio
determined for CH$_{4}$ depends on the data list used and on the H$_{2}$O
mixing ratio. 
For standard quantity of H$_{2}$O ($4.5 \times 10^{-4}$),  the retrieved mixing ratio of methane  is $\sim 10^{-5}$   if we use Nassar and Bernath (2003) 
line lists. Mixing ratios 10-50 times larger are needed for methane if we use PNNL or HITRAN 2008.
The HITRAN 2008 data bank has the advantage of covering the entire spectral range measured by NICMOS, with 
the downside (shared also by the PNNL list)  is that it results from measurements at room temperature, and therefore is quite inadequate to estimate the mixing ratio of methane at the temperatures of interest here. The Nassar and Bernath (2003) data provide a much better fit to our observations in the region where they overlap with HITRAN 2008.
The contribution of CO$_2$ is clearly shown in Fig. \ref{mol}, and its mixing ratio is consistent with the one of H$_{2}$O.
An upper limit of $\sim 10^{-2}$  was found for CO.

\begin{figure}
\includegraphics[width=8cm]{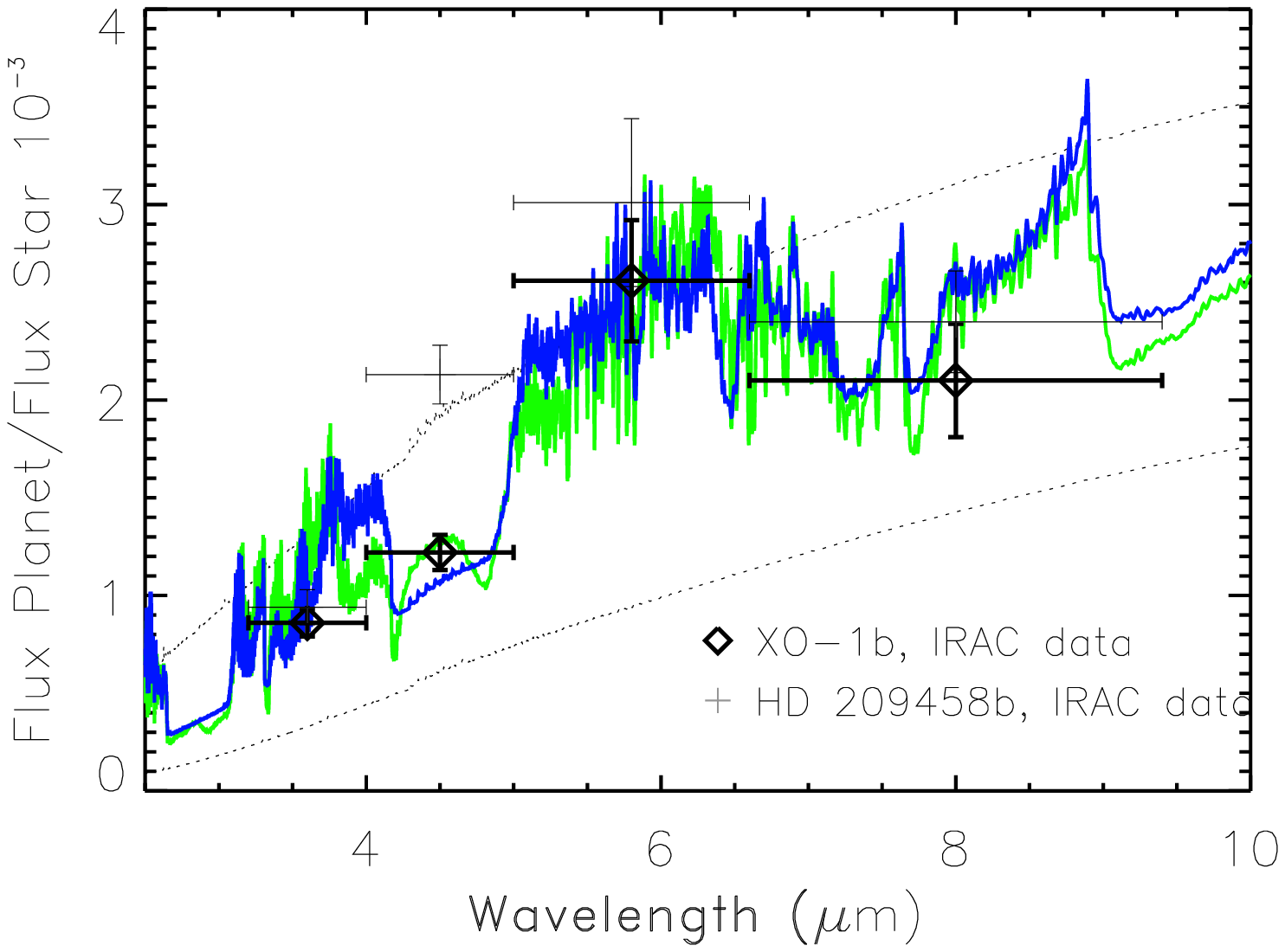}
\includegraphics[width=9cm]{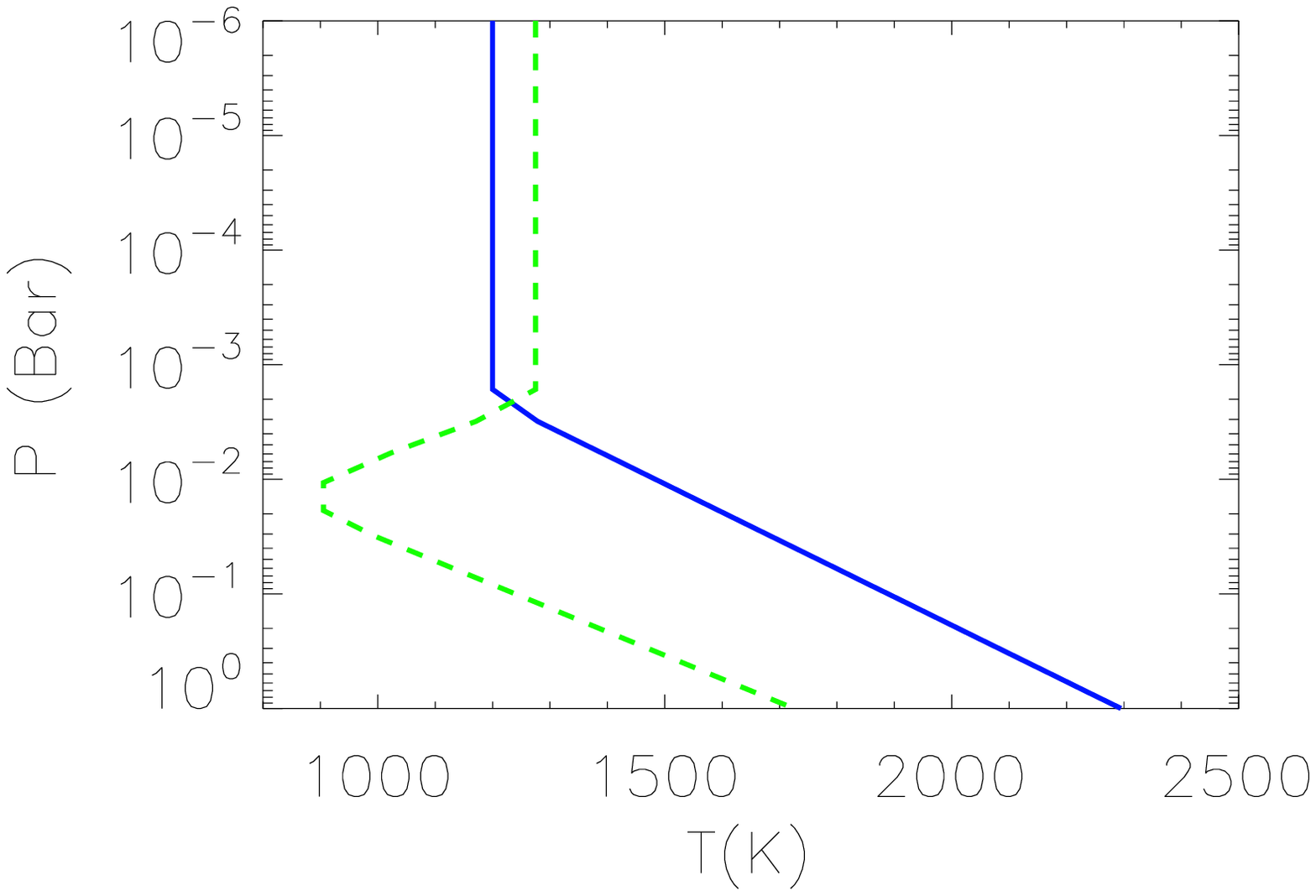}
\caption{Left panel: {\em Spitzer}-IRAC photometric data obtained during the secondary transit of XO-1b (Machalek et al., 2008) and HD 209458b (Knutson et al., 2008). 
The dotted lines show the blackbody curves at 1000 K and 1500 K.
The green and blue line show the spectra simulated for XO-1b using atmospheric models for XO-1b containing H$_{2}$O, CH$_{4}$ and CO$_2$. The spectrum obtained using a $T-P$ profile  with no temperature inversion (see the right-hand panel) and mixing ratios compatible with the night side is plotted in blue, while the one obtained using a $T-P$ profile  with  temperature inversion (see left-hand panel) is plotted in green.
Right panel: two examples of $T-P$ profiles used for the spectral interpretation of the XO-1b photometric data plotted in the left-hand panel. The 
retrieved  mixing ratios for  H$_{2}$O, CH$_{4}$ and CO$_2$
are higher in the case of no inversion.}
\label{xo3}
\end{figure}

As previously mentioned, these observations of XO-1b allow a direct
comparison with HD 209458b. 
Concerning the NICMOS transmission spectra, we find that H$_{2}$O, CH$_{4}$, CO$_2$ and possibly CO explain the spectra for  XO-1b and HD209458b.  
However, a haze component may be present in the HD 209458b atmosphere while  the XO-1b data do not require it.  
For consistency, we also tried to fit the photometric data points obtained by
Machalek et al. (2008) using the secondary transit technique. We have tried different classes of 
$T-P$ profiles (Fig. \ref{xo3}) as we did in the case of  HD 209458b (Swain et al., 2009b).
The assumption that the chemical abundances of the atmospheric components do not change dramatically from the day to the night side needs further confirmation. To maintain mixing ratios of H$_{2}$O, CH$_{4}$ and CO$_2$ consistent to those adopted to fit the terminator spectroscopy data,
a $T-P$ profile relatively hot towards the 1 bar level and with $T$ decreasing with altitude is needed (blue curve in Fig. \ref{xo3}), while classes of $T-P$ profiles with an inversion
(e.g. green curve) need to be combined with smaller mixing ratios to match the IRAC photometric data. 
Given the highly degenerate nature of the problem, we cannot conclude which thermal profile and combination of mixing ratios characterize  the day-side atmosphere.
Broad-band emission photometry data alone cannot provide an accurate retrieval of four molecules and of a $T-P$ profile. Additional spectroscopic data are needed 
to constrain  the number of plausible solutions. 
By comparing the IRAC  emission photometry data for XO-1b with the same datasets for HD 209458b (Knutson et al., 2007), we clearly see that HD 209458b
is slightly hotter, as expected, but 
there is also a hint
that the CO$_2$ and/or CO
abundance may be higher in XO-1b due to the more pronounced absorption in the 4.5 $\mu$m band (Fig. 3).

\section{Conclusions}
We present in this paper the first transmission spectrum of the
hot-Jupiter XO1b, recorded with the HST-NICMOS instrument. The data
can be explained mainly with the combined presence of   H$_2$O, CH$_4$  and CO$_2$  in the atmosphere of the planet, but we suggest
that CO might also be present in XO-1b. The photometric
emission data observed by Machalek et al. (2008) are consistent with the above composition but a variety of $T-P$ profiles and mixing ratios are compatible with the data. 
In particular $T-P$ profiles with no inversion and a hotter temperature at the 1 bar pressure level,
are compatible with the mixing ratios that fit the terminator spectroscopy
data presented here. Other thermal profiles with a stratosphere can fit the day-side observations if we  change the mixing ratios of the atmospheric components 
proposed for the terminator. 
With only four broad-band emission photometry points we cannot retrieve simultaneously
the thermal structure and the molecular composition from the data, a task that requires additional spectroscopic data.

Finally we compare the atmospheres of two hot-Jupiters showing stellar, orbital, and planetary
similarities,  i.e.\ XO1b and HD 209458b, for which we have data obtained with the same instrument and technique. While these data are consistent with a similar atmospheric composition for the two planets, hints of the presence of hazes on HD 209458b and more abundant CO and/or CO$_2$ on XO-1b show a departure from the 
affinity.
 Additional observations of transiting hot-Jupiters,
especially spectroscopic data, will allow a more thorough
classification of this type of planets unknown in our Solar System.

\ack G.T. is supported by a Royal Society University
Research Fellowship. Part of the research described in this paper was performed
at the Jet Propulsion Laboratory, California Institute of Technology, under contract with the National Aeronautics and Space Administration.
Special thanks to Jonathan Tennyson and Peter Bernath for their contribution on molecular data lists.

\end{document}